\documentclass[%
    aps,
    prb,
    reprint,
    superscriptaddress,
    amsfonts,
    amsmath,
    amssymb,
    eprint,
    longbibliography,
]{revtex4-2}

\usepackage{silence}
\usepackage{graphicx}
\usepackage{dcolumn}
\usepackage{bm}
\usepackage{mathtools}
\usepackage{physics}
\usepackage{nicematrix}
\usepackage{stmaryrd}
\usepackage{hyperref}
\usepackage{cleveref}
\Crefname{figure}{Fig.}{Figs.}
\Crefname{equation}{Eq.}{Eqs.}
\usepackage{import}
\usepackage{color}
\usepackage{lipsum}
\usepackage[disable]{todonotes}
\usepackage{verbatim}
\usepackage{tikz}
\usetikzlibrary{arrows.meta}
\usetikzlibrary{backgrounds}
\usetikzlibrary{arrows,decorations.pathmorphing}

\usepackage[caption=false]{subfig}

\usepackage{textgreek}
\usepackage{etoolbox}
\apptocmd{\sloppy}{\hbadness 10000\relax}{}{}

\begin{document}

\title{\texorpdfstring{$\Delta T$}{ΔT}-noise in Multiterminal Hybrid Systems}

\author{Leonardo Pierattelli}
\affiliation{NEST, Istituto Nanoscienze-CNR and Scuola Normale Superiore, Piazza San Silvestro 12, I-56127 Pisa, Italy}
\author{Fabio Taddei}
\affiliation{NEST, Istituto Nanoscienze-CNR and Scuola Normale Superiore, Piazza San Silvestro 12, I-56127 Pisa, Italy}
\author{Alessandro Braggio}
\affiliation{NEST, Istituto Nanoscienze-CNR and Scuola Normale Superiore, Piazza San Silvestro 12, I-56127 Pisa, Italy}

\date{\today}

\begin{abstract}

The study of charge current fluctuations (noise) can give useful insights into the properties of nanoscale systems.
In this work, the peculiar properties of noise in multiterminal hybrid normal-superconducting systems are explored in the thermal out-of-equilibrium regime, i.e., when temperature biases are present ($\Delta T$-noise).
Using the Landauer-B\"uttiker approach, we identify two contributions: background noise and excess noise, analyzing them
when both electrical and thermal biases are applied.
When temperature biases are present, and superconducting terminals are grounded, we find that the first contribution depends not only on the electrical conductance, as the Johnson-Nyquist at equilibrium, but also on a quantity strictly related to the heat conductance. 
This is our first main result.
On the other hand, the second contribution shows, as expected, additional terms originating from the partitioning of currents into different transport channels, including the ones associated with Andreev reflection processes. However, noise induced by the temperature differences unveil also interference terms that cannot be present either in voltage bias or in the absence of any Andreev processes.
Finally, we apply the results obtained to two different specific physical situations.
The first is a generic three-terminal normal-superconductor-normal system, and the second is a device based on spin-resolved co-propagating chiral channels in the integer quantum Hall regime with a superconducting region.
In these example setups, we investigate mainly the shot-noise regimes, when high-voltage or high-temperature biases are applied. We find remarkable differences between the two limits, which ultimately show the different nature of electrically and thermally induced charge current fluctuations.
\end{abstract}

\maketitle

\section{Introduction}

It is a well-known fact that charge current correlations are a valuable source of information in electronic systems~\cite{landauerNoiseSignal1998}, thanks to their capability of probing various effects and observables.
These can range from the charge of current carriers~\cite{schottkyUberSpontaneStromschwankungen1918,de-picciottoDirectObservationFractional1997,saminadayarObservation$e3$1997,ferraroChargeTunnelingFractional2010}, their statistical properties~\cite{de-picciottoDirectObservationFractional1997,saminadayarObservation$e3$1997,safiPartitionNoiseStatistics2001,bartolomeiFractionalStatisticsAnyon2020}, to the thermodynamic quantities of 
devices~\cite{johnsonThermalAgitationElectricity1928,whiteStatusJohnsonNoise1996}, the quantum features in coherent mesoscopic systems~\cite{blanterShotNoiseMesoscopic2000} or even the intrinsic limitations of quantum machine performances~\cite{baratoThermodynamicUncertaintyRelation2015,gingrichDissipationBoundsAll2016,pietzonkaUniversalBoundsCurrent2016,horowitzThermodynamicUncertaintyRelations2020}.
Indeed, thanks to all recent advancements in nanofabrication, the application of the properties of charge current noise will enable more and more promising technologies, for example, making use of interferometric effects in Quantum Hall systems~\cite{jiElectronicMachZehnder2003,beenakkerProposalProductionDetection2003,nederEntanglementDephasingPhase2007,bocquillonElectronQuantumOptics2014}, or for Quantum State Tomography in electronic quantum devices~\cite{samuelssonQuantumStateTomography2006}.
Nonetheless, the same amount of interest is being devoted also in hybrid normal- or Quantum Hall-superconducting devices~\cite{taddeiPositiveCrosscorrelationsInduced2002,vanostaaySpintripletSupercurrentCarried2011,ferraroNonlocalInterferenceHongOuMandel2015,ametSupercurrentQuantumHall2016,houCrossedAndreevEffects2016,golubevCrosscorrelatedShotNoise2019}, mainly thanks to their relevance nowadays in the field of quantum technologies and computing~\cite{clarkeExoticCircuitElements2014,marchegianiNoiseEffectsNonlinear2020}.
By engineering the combination between existing normal platforms with superconductors, one can again investigate the noise properties in order to detect several important effects, e.g. the quasiparticles' nature~\cite{bolechObservingMajoranaBound2007,beenakkerAnnihilationCollidingBogoliubov2014,thupakulaCoherentIncoherentTunneling2022} or the basic features of the systems we are dealing with, including the presence and the type of interactions~\cite{braggioShotNoiseQuantum2003,hofferberthProbingQuantumThermal2008,braggioSuperconductingProximityEffect2011,arracheaSignaturesTripletSuperconductivity2024}.

Historically, the focus has been devoted to the electrically out-of-equilibrium characteristics of current fluctuations.
However, there has recently been a surge in the interest around the thermally out-of-equilibrium ones~\cite{lumbrosoElectronicNoiseDue2018,sivreElectronicHeatFlow2019,larocqueShotNoiseTemperatureBiased2020,hasegawaDelta$T$NoiseKondo2021,popoffScatteringTheoryNonequilibrium2022,zhangDelta$T$NoiseWeak2022,hublerLightEmissionDelta$T$driven2023,shein-lumbrosoDeltaTFlickerNoise2024}, also in Quantum Hall systems~\cite{rechNegativeDelta$T$Noise2020,reboraDelta$T$NoiseFractional2022,iyerColoredDelta$T$Noise2023}.
This is justified by the fact that exploring those regimes can give access to, for example, additional and complementary information on the presence of interactions in a certain system, of the quantum states therein, or of hotspots in quantum devices~\cite{scheerUnexpectedNoiseHot2018,mishraFinitetemperatureQuantumNoise2023,braggioNonlocalThermoelectricDetection2024}.
Although such thermally-induced noise fluctuations are limited by very general bounds~\cite{erikssonGeneralBoundsElectronic2021,tesserChargeSpinHeat2023}, there is room for relevant applications, such as thermometry~\cite{prokudinaLocalThermometryNbSe$_2$2024}, and quasiparticle manipulation and detection~\cite{husseinNonlocalThermoelectricityCooperpair2019,golubevCooperPairSplitting2023,smirnovMajoranaDifferentialShot2023,mohapatraProbingYuShibaRusinovState2024}.
Again, the study of this kind of noise in hybrid normal-superconducting systems is of primary importance for future applications of thermally biased systems~\cite{giazottoFerromagneticInsulatorBasedSuperconductingJunctions2015,guarcelloJosephsonThresholdCalorimeter2019,paolucciHighlySensitiveBroadband2023} but, to the best of our knowledge, until now the analysis has been devoted mainly to the study of the noise characteristics of temperature-biased NS junctions~\cite{zhitlukhinaElectronicNoiseGenerated2020,golubevCooperPairSplitting2023,mohapatraAndreevReflectionMediated2024}.

In this paper, we investigate the general properties of noise under stationary out-of-equilibrium configurations of \textit{multiterminal} nanoscale hybrid superconducting systems, particularly focusing on the effects determined by temperature differences ($\Delta T$-noise).
In the first part, we review the Landauer-Büttiker theory of coherent electron transport in the presence of grounded superconductors for a general multiterminal setup~\cite{anantramCurrentFluctuationsMesoscopic1996,moskalets2011scattering}, while recalling the definitions of the Onsager matrix elements which are useful to write the noise expressions in a compact form.
We mainly concentrate on the charge current self- and cross-correlations at zero-frequency, where we identify two contributions of different nature: the background (or thermal) noise and the excess (or partition) noise.
We analyze their main features and behaviour with respect to the applied external biases.
Regarding the first contribution to noise, we discuss a fundamental distinction between the normal and superconducting cases, which emerges in the presence of temperature differences: our main finding is that the background noise part of self-correlations, at any given terminal, contains terms strictly related to the heat conductances, relative to the other terminals.
These are $\Delta T$-noise contributions, which effectively measure the thermal noise contributions determined by all other terminals and reduce to the standard Johnson-Nyquist noise at thermal equilibrium.
On the other hand, the
excess noise in multiterminal hybrid systems exhibits a more conventional behaviour.
Since the excess noise is the sum of all the partition noises that originated in the system,
in our case, we have additional terms corresponding to the Andreev processes, even when they are determined by temperature differences only, and we obtain the known result that such terms can be positive in hybrid systems.
In particular, a full characterization of this contribution is done in the case of an energy-independent scattering matrix.
We then study the \textit{electrical} shot noise, i.e. the excess noise produced by applying strong voltage differences in the terminals, and also what we call \textit{thermal} shot noise, which is the excess noise produced by applying strong temperature differences instead.

Finally, we apply the approach developed to two examples that are sufficiently general to show the important facts of the general analysis presented before. The first system consists of a three-terminal normal-superconductor-normal (NSN) junction, and the second one is a multiterminal device based on the integer Quantum Hall effect with a pair of co-propagating spin-resolved chiral edge states containing a proximate superconducting region.
We mainly analyze the properties of the excess noise in various regimes, highlighting the role of thermally induced current fluctuations and contrasting them with the more conventional electrical ones.
In particular, we start from a consideration developed in the past literature, and that is, noise in hybrid systems can be used to probe not only the \textit{electrical} conductance of a device, e.g. via Johnson noise measurements, but also the \textit{heat} conductance~\cite{akhmerovQuantizedConductanceMajorana2011,bubisThermalConductanceNonequilibrium2021,denisovChargeneutralNonlocalResponse2021,denisovHeatModeExcitationProximity2022}.
In this work, we recall that this is a peculiarity of the electrical shot noise of self-correlators, contrasting it with the \textit{thermal} shot noise results, never discussed before, where a much clearer dependence on heat conductance is recognizable.
This observation highlights substantial differences in the nature of electrically induced and thermally induced charge current fluctuations in hybrid systems, which, in conclusion, can be noticed by comparing the electrical and thermal shot noise of cross-correlators.

\section{Current Correlations in Hybrid Multiterminal Systems}

In the following, we wish to present the general theory of the current correlations for a hybrid superconducting multiterminal system out of equilibrium. 
We will review the general formulae in the scattering formalism for finite voltage and temperature biases. 
We first recall the formulae of the average currents, and we concentrate on the results for the cross-correlations. 
The approach will be completely general such that we could easily apply those formulae from a few terminals to chiral quantum Hall edge states with or without superconducting terminals.  

\subsection{Hamiltonian and Scattering Matrix}
We consider the general case of a hybrid superconducting system, taking into account the various quantum numbers that characterize its states, including spin.
The system consists of a generic scattering region comprising both normal and superconducting regions connected, through leads, to $M$ normal contacts. The latter are at local thermodynamical equilibrium, each characterized by a temperature $T_i$ and an electrochemical potential $\mu_i$, with $i=1,..., M$. Each lead allows a number $N_i$ of transport channels, possibly dependent on the energy. All superconducting regions are assumed to be at the same electrochemical potential $\mu_s$.
Such a setup can be conveniently described using the Bogoliubov-de Gennes (BdG) Hamiltonian approach \cite{altlandNonstandardSymmetryClasses1997}:
\begin{equation}
    H = \begin{pmatrix}
        H_0           &   \Delta      \\
        -\Delta^*    &   -H_0^T
    \end{pmatrix},
    \label{eq:bdg}
\end{equation}
where $H_0$ and $\Delta$ are $2\times 2$ matrices in the spin space to fully account for the spin degree of freedom.
In particular, $H_0$ accounts for the particle degree of freedom, $-H_0^T$ for the hole degree of freedom, and $\Delta$ is the superconducting order parameter that couples particles and holes \cite{lambertPhasecoherentTransportHybrid1998,jacquodOnsagerRelationsCoupled2012}.
To write the BdG Hamiltonian we have used the basis of Ref.~\cite{altlandNonstandardSymmetryClasses1997}, i.e. 
\begin{equation}
    \mathbf{c}(E) = \begin{pmatrix}
        c_{ieq\uparrow}(E) ,  &
        c_{ieq\downarrow}(E),   &
        c_{ihq\uparrow}(E) ,  &
        c_{ihq\downarrow}(E)
    \end{pmatrix}^T,
    \label{eq:spinorial}
\end{equation}
where $c_{i\alpha q\sigma}(E)$ is the annihilation operator for an $\alpha$-like particle ($\alpha=e\equiv +1$ for electrons and $\alpha=h\equiv -1$ for holes) with spin $\sigma=\uparrow,\downarrow$ in lead $i$ with channel index $q$.
We measure the excitation energies $E$ with respect to the \textit{common} electrochemical potential $\mu_s$ of all the superconductors of the system.
Using the scattering theory for normal-superconducting devices \cite{anantramCurrentFluctuationsMesoscopic1996,blanterShotNoiseMesoscopic2000}, we define the scattering matrix $s(E)$ which connects the scattering states' creation operators $\mathbf{a}(E)$, \textit{incoming} to the scatterer, with the \textit{outgoing} ones $\mathbf{b}(E)$ via the equation $\mathbf{b}(E) = \mathbf{s}(E) \mathbf{a}(E)$, assuming \textit{elastic} scattering processes.
The scattering matrix has the following structure
($\alpha,\beta\in\{e,h\}$):
\begin{gather}
    \mathbf{s}(E)\equiv s^{\alpha\beta}_{ij}(E) = \begin{pmatrix}
        s^{\alpha\beta}_{ij,11}(E)    &   \hdots  &  s^{\alpha\beta}_{ij,1 N_j}(E)   \\
        \vdots                  &   \ddots  &   \vdots  \\
        s^{\alpha\beta}_{ij,N_i 1}(E)    &   \hdots  &   s^{\alpha\beta}_{ij,N_i N_j}(E)
    \end{pmatrix},
    \label{eq:smatchan}
\end{gather}
where each element describes the probability amplitude for a particle of type $\beta$ from lead $j$ to be scattered into lead $i$ as a particle of type $\alpha$, between all pairs of open channels in the leads.
The Fermi distribution in the contact $i$ with electrochemical potential $\mu_i$ and temperature $T_i$, for $\alpha$-like particles, is defined as $f^\alpha_{i}(E) = \left\{ \exp[\left(E-\text{sgn}(\alpha)e V_{i}\right)/k_B T_i]+1 \right\}^{-1}$, which satisfies the particle-hole symmetry (PHS) property $f^{h}_{i}(E) = 1-f^e_{i}(-E)$.
In addition, the voltage biases are measured from the superconducting electrochemical potential, i.e. $e V_{i}=\mu_i-\mu_s$, and in the following we fix $\mu_s=0$.

\subsection{Currents}
Using standard scattering theory~\cite{anantramCurrentFluctuationsMesoscopic1996,benentiFundamentalAspectsSteadystate2017}, the average electrical and energy currents can be written in the following form (assuming they are \textit{positive} when \textit{entering} the scattering region):
\begin{equation}
    \renewcommand{\arraystretch}{1.2}
    \begin{bmatrix}
         J^{\rm C}_i
         \\
         J^{\rm U}_i
    \end{bmatrix}
    =
    \int
    \frac{dE}{2h}
    \sum_{k=1}^{M}
    \begin{bmatrix}
         e 
         \ell^+_{ik}(E) 
         \\
         E \ell^-_{ik}(E)  
    \end{bmatrix}
    \! \! 
    \left[
        2f^e_{k}(E)
        \! 
        -
        \! \! 
        f^e_{i}(E)
        \!
        - 
        \! \! 
        f^h_{i}(E)
    \right]
    .
    \label{eq:curravg}
\end{equation}
where the integration extrema are $(-\infty,+\infty)$.
Additionally, the \textit{heat} current can be easily computed from $J^{\rm H}_i=J^{\rm U}_i- V_{i} J^{\rm C}_i$.
The ``transmission" functions $\ell^\pm_{ik}$ are defined as follows:
\begin{equation}
    \ell^{\pm}_{ik}(E) =
        N_i\delta_{ik} \! - \!
        \Tr[
            s^{ee\dagger}_{ik}(E)s^{ee}_{ik}(E)
        ] \! \pm \! \Tr[
            s^{he\dagger}_{ik}(E)s^{he}_{ik}(E)
        ] \! .
    \label{eq:ellpm}
\end{equation}
The traces $\Tr[...]$ terms are the normal ($ee$) and Andreev ($he$) transmission coefficients between contacts $i$ and $k$, and we notice that $\ell^+$ and $\ell^-$ differ just by a single minus sign in front of the Andreev term.
The difference accounts for the fact that the holes carry an opposite charge with respect to the particles but the same amount of energy, and indeed, electrons that undergo Andreev processes contribute oppositely to the electrical current in the arrival contact.
As a side note, we remark the fact that in a purely normal device, no Andreev processes occur, i.e. $\Tr[s^{he\dagger}_{ik}(E)s^{he}_{ik}(E)]=0$, hence $\ell_{ik}^+$ and $\ell_{ik}^-$ would be identical. 
As a consequence of their definitions, the following properties hold for $\ell_{ik}^\pm$:
\begin{gather}
    \sum_{k=1}^M\ell^-_{ik}(E) = 0 \label{eq:ellminuscons}
    \\
        \sum_{k=1}^M \ell^+_{ik}(E) 
        \!
        =
        \!
        \sum_{k=1}^M
            \Tr[
                s^{eh\dagger}_{ik}(E)
                s^{eh}_{ik}(E)
            ]
            \!
            +
            \!
            \Tr[
                s^{he\dagger}_{ik}(E)
                s^{he}_{ik}(E)
            ],
    \label{eq:unitarity}
\end{gather}
Note that \Cref{eq:ellminuscons} expresses the energy conservation, while
\Cref{eq:unitarity} expresses the fact that, due to the Andreev reflections, the charge is not conserved across the normal contacts at least when the superconducting regions in the scatterer are grounded \footnote{On the other hand, if the superconductors were floating, we would also get $\sum_k \ell^-_{ik}(E)=0$ in the stationary conditions.}.
The right-hand-side of \Cref{eq:unitarity} is sometimes called excess conductance~\cite{lambertPhasecoherentTransportHybrid1998,vanweesExcessConductanceSuperconductorsemiconductor1992}.

In the linear response regime, where small voltage $e\delta V_{k} \ll k_BT$ and temperature biases $\delta T_{ki}=T_k-T_i\ll T$ are assumed with $T$ is the equilibrium temperature. The currents can be expressed in terms of multiterminal Onsager matrices $\mathbf{L}_{ik}$ as
\cite{jacquodOnsagerRelationsCoupled2012,benentiFundamentalAspectsSteadystate2017}
\begin{equation}
    \begin{aligned}
        \renewcommand{\arraystretch}{1.2}
        \begin{pmatrix}
             J^{\rm C}_i \\
             J^{\rm H}_i
        \end{pmatrix} & =
        \sum_{k=1}^M
        \mathbf{L}_{ik} \begin{pmatrix}
            \delta V_{k}/T   \\
            \delta T_{ki}/T^2
        \end{pmatrix} ,
        \label{eq:onsager}
    \end{aligned}
\end{equation}
where
\begin{equation}
    \begin{aligned}
    \mathbf{L}_{ik}= \begin{pmatrix}
            L^{\rm{CC}}_{ik} &   L^{\rm{CH}}_{ik} \\
            L^{\rm{HC}}_{ik} &   L^{\rm{HH}}_{ik}
        \end{pmatrix}
\end{aligned}.
\end{equation}
where we have also
introduced the affinities $\delta V_{k}/T$ and $\delta T_{ki}/T^2$. From \Cref{eq:curravg} we can write the Onsager matrix elements as \cite{claughtonThermoelectricPropertiesMesoscopic1996}
\begin{equation}
    \renewcommand{\arraystretch}{1.5}
        \mathbf{L}_{ik}
    = \int \frac{dE}{h}
        \left[
            - \frac{\partial f(E)}{\partial E}
        \right]
        \begin{pmatrix}
            e^2 T \ell^+_{ik}(E) 
            &
            eTE\ell^+_{ik}(E) \\
            eTE\ell^-_{ik}(E) 
            &
            TE^2\ell^-_{ik}(E) \\
        \end{pmatrix},
        \label{eq:conductmatr}
\end{equation}
where $f(E) = \left[ \exp(E/k_B T)+1 \right]^{-1}$ is the electron equilibrium distribution at temperature $T$. 
When the energy dependencies can be neglected, i.e. $\ell^{\pm}_{ik}(E)\equiv \ell^{\pm}_{ik} $, we find the usual expressions for the charge conductances $L^{\rm{CC}}_{ik}=G_0 T\ell_{ik}^+$ (with $G_0=e^2/h$ being the conductance quantum) and for the heat conductances $L^{\rm{HH}}_{ik}=(\pi^2/3) (k_B^2/e^2)G_0 T \ell^-_{ik}$ \footnote{Note that, in the linear regime, the heat is conserved as direct consequence of the energy conservation of \Cref{eq:ellminuscons} since the terms which break the heat conservation are at least $\mathcal{O}(\delta V_i^2)$.}. 
In the energy-independent case, the off-diagonal terms (proportional to the Seebeck and Peltier coefficients) are null, since the integrand is odd in energy and, therefore, there are no thermoelectrical effects.
We also notice that in this same limit in the normal case, we immediately recover the Wiedemann-Franz law $L^{\rm{HH}}_{ik}/L^{\rm{CC}}_{ik}=(\pi^2/3) (k_B^2/e^2)T$.
However, this is known to be violated when superconductors are present, and in fact, in general, one has $\ell^-_{ik}/\ell^+_{ik}\neq 1$.

\subsection{Correlations}
Using the scattering approach we can also compute the charge current correlations. In particular, we are interested in the symmetrized definition of the correlators, or current-noise power spectrum in stationary conditions~\cite{blanterShotNoiseMesoscopic2000}:
\begin{equation}
    S_{ij}(\omega) = \int_{-\infty}^{+\infty} dt e^{i\omega t}
    \langle
        \delta \hat J^{\rm C}_i(t) \delta \hat J^{\rm C}_j(0)
        + \delta \hat J^{\rm C}_j(0) \delta \hat J^{\rm C}_i(t)
    \rangle,
    \label{eq:corrtime}
\end{equation}
where
$\delta \hat J^{\rm C}_i(t)=\hat{J}^{\rm C}_i(t)- J_i^{\rm C} $~\footnote{Note that in the definition of the Fourier transform a prefactor $2$ is added because of the widely used engineering convention to define the spectrum only for positive frequencies.}.
We will concentrate the discussion on investigating only the zero-frequency limit, which is also more easily measurable.
We can then divide the noise expressions in two contributions $S_{ij}(0) = \Bar{S}_{ij} + \tilde{S}_{ij}$ as, until now, has been done previously for normal systems~\cite{moskalets2011scattering,erikssonGeneralBoundsElectronic2021,popoffScatteringTheoryNonequilibrium2022,tesserChargeSpinHeat2023}, based on the dependence over various Fermi function distributions' expressions.
The first term is what we will call background noise, containing only expressions of the type $F^e_{i}(E)\equiv 2f^e_{i}(E)\left[ 1-f^e_{i}(E) \right]$~\footnote{The equivalent quantities from holes' distributions, $F^h_{i}(E)$, are not present simply because we usually apply the PHS to further simplify the expressions.}:
\begin{equation}
    \begin{multlined}
    \Bar{S}_{ij} = G_0 \int dE
    \Bigg[
        F^e_{j}(E) \ell^+_{ij}(E)
        + F^e_{i}(E) \ell^+_{ji}(E) \\
        - \delta_{ij} \sum_{k}^{M} F^e_{k}(E) \ell^-_{ik}(E)
    \Bigg],
    \end{multlined} 
    \label{eq:back} 
\end{equation}
where the functions $\ell^-_{ik}$ are present only in the self-correlators ($i=j$).
At equilibrium, i.e. $T_k=T,V_k=0\, \forall k$,
this term contains the so-called Johnson-Nyquist thermal noise \cite{johnsonThermalAgitationElectricity1928} and it is the only surviving term of the noise.
Notably, this is also the only term that survives when the values of the biases are small, i.e. $eV,k_B\Delta T\ll k_B T$. In this limit, intriguingly, this term can be considered   
the leading term of noise in linear regime (see also later discussions).
However, when the system goes well beyond quasi-equilibrium (namely, well beyond linear regime), one needs to consider the contribution also from the second term of the noise, $\tilde{S}_{ij}$, which can be called excess noise  \cite{lumbrosoElectronicNoiseDue2018} and it contains only squared differences of Fermi functions $D_{k\gamma l\delta}(E)\equiv [f^{\gamma}_{k}(E)-f^\delta_{l}(E)]^2$:
\begin{equation}
    \begin{multlined}
        \tilde{S}_{ij} = G_0 \int dE
        \sum_{\alpha}
        \text{sgn}(\alpha)
        \sum_{\substack{kl\\\gamma\delta}}
        D_{k\gamma l\delta}(E) \\
        \times \Tr[
            s^{e\gamma\dagger}_{ik}(E)
            s^{e\delta}_{il}(E)
            s^{\alpha\delta\dagger}_{jl}(E)
            s^{\alpha\gamma}_{jk}(E)
        ],
    \end{multlined}
    \label{eq:excess}
    \end{equation}    
where, from now on, the Greek indices $\alpha,\gamma,\delta$ range over the particle types $e,h$ and the Latin ones $k,l$ over the contact labels from $1$ to $M$.
At thermal equilibrium ($T_k=T\, \forall k$), these terms contain the expected Schottky current noise contributions, which usually emerge at high electrical biases.
Moreover, these terms are proportional to a fourth-order contribution of the scattering matrix, thus clearly encoding the effect of the partition noise \cite{blanterShotNoiseMesoscopic2000}.

In the following, we will study the behaviour of the excess noise under electrical and thermal biases or both.
However, in the excess noise, it is important to study the aforementioned large voltage bias limit and, in our case, to compare it to the large temperature bias limit.
For the sake of simplicity, we call these limits \textit{electrical} and \textit{thermal} shot noise, respectively, and in both limits the excess noise is almost linear in the bias applied.

It could be possible to generalize the previous results also to finite frequency noise.
However, in such cases, the simple scattering approach computation neglects effects due to the displacement currents, which are, in the end, interaction effects which cannot be easily described with a pure scattering approach~\cite{blanterShotNoiseMesoscopic2000}. 
Furthermore, the measurement of frequency-dependent noise is usually much more complex than that for zero-frequency noises, also involving issues connected to the precise quantum measurement prescription~\cite{gabelliHighFrequencyDynamics2009,farleyNoiseDynamicsQuantum2023}.
For these reasons, we do not discuss frequency-dependent effects any further.
Instead, we will concentrate on the structure of the background and the excess noise terms.

\subsubsection{Background Noise}

To better analyze the background noise's structure it is convenient to first investigate it in the limit of energy-independent scattering matrix, i.e. $s^{\alpha\beta}_{ij}(E) \equiv s^{\alpha\beta}_{ij}$, and consequently $\ell^{\pm}_{ik}(E)\equiv \ell^{\pm}_{ik}$.
In general, we keep all the terminals at different temperatures $T_i$ and different voltages $V_i$.
However, in the limit of energy-independent scattering matrices, the background terms are completely independent of the different biases $V_i$ and one finds:
\begin{equation}
    \Bar{S}_{ij} =
    2 k_B G_0
    \left[
        T_j \ell^+_{ij}
        + T_i \ell^+_{ji}
        - \delta_{ij} \sum_k T_k \ell^-_{ik}
    \right].
    \label{eq:backlinenindell}
\end{equation}
The structure of the different terms of the background noise clearly resembles a sum of standard Johnson-Nyquist noise contributions generated by each terminal, with the first two weighted with the transmission coefficients $\ell^+_{ij}$ and the others in the sum with $\ell^-_{ik}$.
At equilibrium ($T_k=T,V_k=0\, \forall k$), the first two terms give exactly the Johnson-Nyquist noise \cite{johnsonThermalAgitationElectricity1928}, while the others cancel out, a direct consequence of the energy conservation \Cref{eq:ellminuscons}.
These last terms appear only in the self-correlations ($i=j$) in the presence of thermal biases, and they are proportional to the heat conductances $L^{\rm{HH}}_{ik}$.

To better appreciate the physical meaning of this result one needs firstly to address why the third term of \Cref{eq:backlinenindell} is relevant only for the self-correlators.
This can be justified in this way: in every lead $i$, the current flowing in it will fluctuate by $\delta J^{\rm C}_i$. 
The fluctuation in the $i$-th terminal can be thought of as being composed of different independent contributions $\delta J^\alpha_{ik}$, each coming from contact $k$ and from different quasiparticles' channels $\alpha=e,h$, i.e. $\delta J^{\rm C}_i = \sum_{k,\alpha} \delta J^\alpha_{ik}$.
Using this picture, in cross-correlations' background noise $\Bar{S}_{ij}$ (that is when $i\neq j$) only the correlator $\sum_{\alpha}\langle \delta J^\alpha_{ij} \delta J^\alpha_{ji} \rangle$ remains and all other terms cancel out since all other fluctuation components coming from different channels are independent of each other~\footnote{This is the usual assumption for a multiterminal setup where each lead operates as an independent current source.}.
On the contrary, in self-correlation background noises $\Bar{S}_{ii}$ there remains the sum over all the channels $k$, $\sum_{k,\alpha}\langle \delta J^\alpha_{ik} \delta J^\alpha_{ik} \rangle$. 
Moreover, such terms must be all positive for all contacts $k$ and particle types $\alpha$, since they are quadratic.
Indeed, we can rewrite \Cref{eq:backlinenindell} as
\begin{equation}
    \frac{\Bar{S}_{ii}}{2k_BG_0}
    \! \!
    =
    \!
        2 T_i \ell^+_{ii}
        + \sum_{k}
        (T_k-T_i)
        \!
        \left(
            \Tr[
                s^{ee\dagger}_{ik}s^{ee}_{ik}
            ] 
            \! \!
            + 
            \!
            \Tr[
                s^{he\dagger}_{ik}s^{he}_{ik}
            ]
        \right)
        \!
        ,
\end{equation}
where we used the unitarity property in \Cref{eq:ellminuscons} and the definition of $\ell^\pm_{ik}$ from \Cref{eq:ellpm}.
Written in this form, it is clear the physical meaning of every term in the expression: the first is just the intrinsic Johnson-Nyquist noise, while the summation is the thermal noise transferred from all other contacts towards the considered terminal.
Note that in both cases, whether the particles are transmitted or they undergo an Andreev reflection, their contributions are always positive irrespective of the charge sign since the term considered is, in the end, a self-correlators' contribution~\footnote{The reader should note that the apparent negative sign in the sum of this form is fully compensated by the Johnson-Nyquist term such that for any temperature the self-correlator is always positive.}.
This generalizes the case of normal systems since, in such cases, it is only the electrical conductance, via $\ell^+$, that determines all the Johnson-Nyquist noise-like contributions.
However, in the presence of grounded superconductors and thermal biases, the first correction to self-correlations' Johnson-Nyquist noise is proportional to the heat conductance instead of the electrical one.
Indeed, the noise contribution to self-correlators from the other terminals is mediated by the excitation transport (heat transport) properties, irrespective of the quasi-particle nature.

\subsubsection{Excess Noise}

\begin{figure}
    \subfloat[\label{fig:excessdv}]{%
        \includegraphics[width=0.45\textwidth]%
        {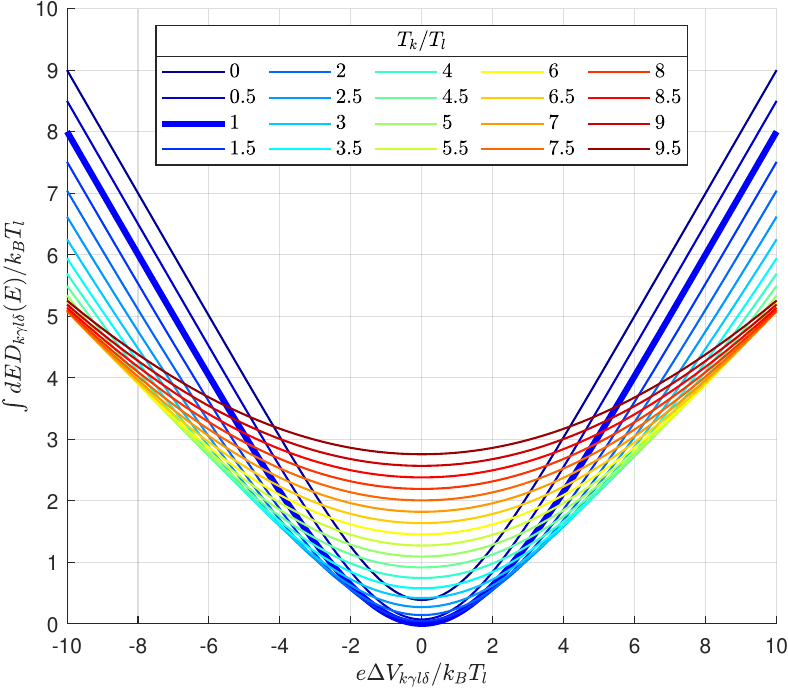}%
    }
    
    \subfloat[\label{fig:excessdt}]{%
        \includegraphics[width=0.45\textwidth]%
        {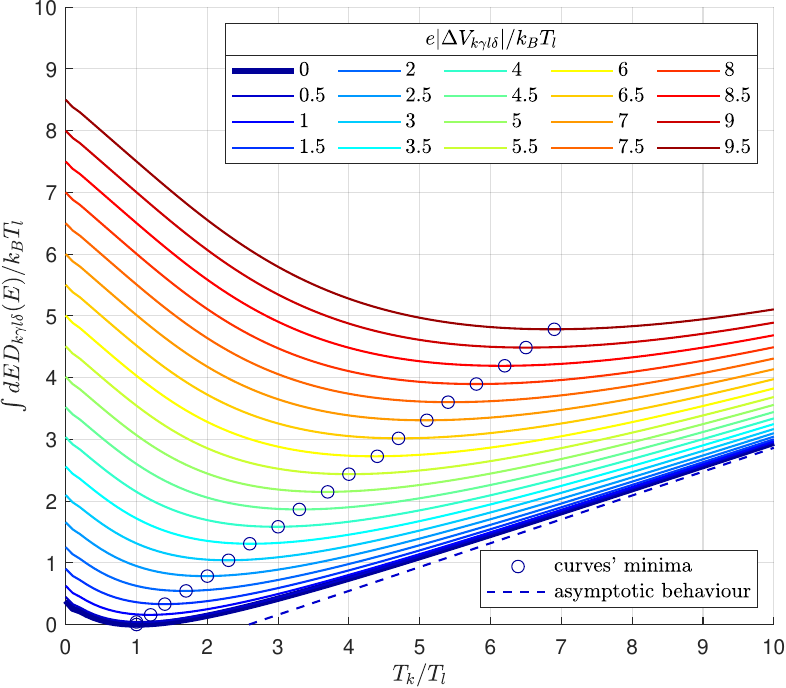}%
    }%
    \caption{Plots of the integrals $\int dE D_{k\gamma l\delta}(E) $ entering the excess noise sum in \Cref{eq:excess-enind}.
    Choosing $k_B T_l$ as the reference energy scale for all quantities,
   panel
    (a) shows the integral's behaviour as a function of $e\Delta V_{k\gamma l\delta}$ (as defined in the text) for different fixed values of $T_k$; panel (b) shows the integral's behaviour as a function of $T_k$ for different fixed values of $e\Delta V_{k\gamma l\delta}$. The dashed line represents the asymptotic behaviour of the curves, with slope $\eta = 2\ln 2 - 1$. The open circles identify the position of the minima of each curve.
    }
    \label{fig:excessplots}
\end{figure}

As stated before, the excess noise defined in \Cref{eq:excess} is the sum of all contributions to noise due to the partitioning processes associated to the scattering of quasiparticles, whatever their type.
More specifically, each contribution in \Cref{eq:excess} is associated to the partitioning of the currents flowing from contacts $k$ and $l$, made up of particles of type $\gamma$ and $\delta$ respectively, into the two currents going into contacts $i$ and $j$ as particles of type $e$ and $\alpha$.

To show the general properties of excess noise, we resort to the same simplifying energy-independent scattering matrix assumption.
In such case, we can write \Cref{eq:excess} as
\begin{equation}
        \tilde{S}_{ij} = G_0
        \sum_{
            \substack{kl \\
            \alpha\gamma\delta}
        }
        \text{sgn}(\alpha)
        \Tr[
            s^{e\gamma\dagger}_{ik}
            s^{e\delta}_{il}
            s^{\alpha\delta\dagger}_{jl}
            s^{\alpha\gamma}_{jk}
        ] \int dE
        D_{k\gamma l\delta}(E),
    \label{eq:excess-enind}
\end{equation}
where all contacts' voltage and temperature dependences are contained in the integrals $\int dE D_{k\gamma l\delta}(E) $, weighted by the partition factor $\left\{\text{sgn}(\alpha) \Tr[ s^{e\gamma\dagger}_{ik} s^{e\delta}_{il} s^{\alpha\delta\dagger}_{jl} s^{\alpha\gamma}_{jk} ]\right\}$.
It can be verified that the integral in \Cref{eq:excess-enind} is symmetrical 
with respect to contact's label exchange between $k$ and $l$, or to the particle label exchange between $\gamma$ and $\delta$.
In fact, what matters are the relative voltages and relative temperatures between different combinations of quasiparticle channels \footnote{It is easy to show, from the form of the quadratic Fermi function difference $D_{k\gamma l\delta}(E)$, that the integral is invariant under exchange of latin \emph{and} greek indices.
However, using the particle-hole symmetry of the Fermi functions $f_i^{\alpha}(E)=1-f_i^{\bar{\alpha}}(-E)$ with $\bar{\alpha}=-\alpha$ and the fact that in the integral the energy is a mute integration variable, in conclusion it can be shown that, in general, is invariant under exchange of latin \emph{or} greek indices.}.
We also recall the fact that the excess noise is negligible in the linear regime: the first non-vanishing terms are of second order in the electrical and thermal biases, and indeed the integral could be approximated at the lowest order as
\begin{equation}
    \int dE
    D_{k\gamma l\delta}(E)
    \approx 
    \frac{
        (e\Delta V_{k\gamma l\delta})^2
        }{
        6 k_B T
    }
     +
    \frac{
        ( \pi^2  - 6)(k_b\Delta T_{kl})^2
        }{
        18 k_B T
    },
\end{equation}
where we denoted the generalized differences between the electrochemical potentials of the two Fermi functions as $e\Delta V_{k\gamma l\delta}\equiv [\text{sgn}(\gamma)e V_{k} - \text{sgn}(\delta)e V_{l}]$, and the temperature difference as $\Delta T_{kl}=T_k-T_l$. It is important to note that when superconductors are present one needs to include also mixed terms where $\gamma\neq\delta$; this implies that $e\Delta V_{k\gamma l\delta}$ can be \emph{sums} of biases other than \emph{differences}.
In other words, we see that this compact formalism using these \emph{generalized} voltage differences can easily account for the case where the partition noise is generated by Andreev reflection processes, and in those cases, the simple voltage differences are substituted by the sum of the voltages with respect to the reference superconducting terminal.

Given these observations, it is useful to discuss the behaviour, under generic biases, of the squared Fermi function difference's integral $\int dE D_{k\gamma l\delta}(E)$ in the excess noise's sum of \Cref{eq:excess-enind}, which weighs each partition factor.
We could go well beyond the small perturbation expansion just discussed before for the excess noise as a function of the biases $e\Delta V_{k\gamma l\delta}$ and/or temperatures $\Delta T_{kl}$. However, since we wish mainly to discuss the expected effects induced by the finite temperature differences it is convenient to measure the temperatures and the energies with respect to the temperature of a reference terminal kept at fixed temperature $T_l$. This latter choice
is a sensible one since allows us to make naturally high and low temperature (ratio) limits.
At the same time, it can fit well the typical experimental situation, e.g. where the temperature in one of the two relevant contacts is kept fixed so that it can be used as a reference for other quantities.

Integration can be performed to evaluate the integral weighting factor in \Cref{eq:excess-enind}, whose results are shown in \Cref{fig:excessplots}. 
Essentially, in these plots, we represent the typical behaviour of the weighting factor in terms of the biases and temperature differences measured in units of the thermal energy $k_B T_l$.
In \Cref{fig:excessdv} it is shown the typical behaviour of the excess noise, as a function of the generalized voltage differences $\Delta V_{k\gamma l\delta}$
keeping fixed the temperature ratio $T_k/T_l$. 
The behaviour is nonlinear when such voltage is too low, but it reduces to the usual linear dependence as soon as $e\Delta V_{k\gamma l\delta}\gg k_B T_k, k_B T_l$, i.e. when the shot-noise limit is reached.
Furthermore, we see how linearity kicks in sooner in voltage when the temperature is lowest, while the curves get smeared, and the explored noise range reduces when the overall temperature in the contacts increases.
However, excess noise could be zero only at equal voltages and identical temperatures, i.e. in the middle of the thicker blue line; otherwise, the minimum of the noise curves is always greater than zero by an amount that is also named noise \textit{thermal floor}.

Indeed, by showing the numerical integration results as a function of $T_k$ while keeping $e\Delta V_{k\gamma l\delta}$ fixed, we obtain instead the curves of \Cref{fig:excessdt}.
The first thing one can notice is the fact that the excess noise increases when temperature differences are present, even in the case of electrical equilibrium (corresponding to the thick blue line, i.e. to the $\Delta T$-noise curve~\cite{lumbrosoElectronicNoiseDue2018}).
In fact in the latter case, for the energy-independent scattering matrix limit that we are showing right now, the average charge currents are zero even in the presence of temperature differences (since we do not have any thermoelectrical effects).
Nonetheless, currents become more noisy as temperatures increase, showing an effect not only in the background noise contribution (see \Cref{eq:backlinenindell}), but also in the excess one, as shown in this section.
However this latter term, unlike the former, is not due to the \textit{transmission} of the charge current thermal fluctuations, but to their \textit{partitioning} across the device.

Having chosen $k_B T_l$ to be the energy reference scale of the system, the plots are not symmetrical around $T_k/T_l=1$.
Moreover, as we said before, going towards both small and large values of $T_k/T_l$, the excess noise increases and its behaviour becomes more linear.
In such limits, the integral's expression can be simplified into
    \begin{align}
        &\lim_{T_k/T_l \to 0^+} \! \!
            \frac{
                \int dE D_{k\gamma l\delta}(E)
            }{ k_B T_l }
        =
        -\frac{T_k}{T_l}
        + \ln \!
        \left[
            \frac{4}{\mathrm{e}}
            \! \cosh^2 \! \!
            \left(
            \!
                \frac{e \Delta V_{k\gamma l\delta}}{2k_B T_l}
            \!
            \right)
        \!
        \right] \! \! , \\
        &\lim_{T_k/T_l \to +\infty} \! \!
        \frac{
            \int dE D_{k\gamma l\delta}(E)
            }{ k_B T_l }
        =
        \eta \frac{T_k}{T_l},
        \label{eq:thermalshot}
    \end{align}
where $\mathrm{e}$ is Euler's number and $\eta\equiv 2\ln{2} - 1$ is a prefactor in the asymptotic behaviour that has also been previously discussed in the literature \cite{larocqueShotNoiseTemperatureBiased2020} and is potentially connected to the Landauer limit $2\ln(2) k_B T_l$ for the minimum heat generated by the erasure of one bit of information \cite{landauerIrreversibilityHeatGeneration1961}.
When noise approaches these regimes, we say we have reached a \textit{thermal} shot-noise regime, in contrast with the shot noise behaviour that we get in the presence of strong voltage biases.
In the following, we will name the latter as \textit{electrical} shot-noise regime for the sake of clarity.

Another thing to notice is the fact that when the voltage bias between the contacts is increased, the curves get enhanced
and their minima sit at a higher noise value (marked with open dots, which in this case can be called noise \textit{voltage floor}, analogously to before): more quasiparticle excitations are flowing into the system, hence more partitioning processes must occur in the system, and more noise is produced associated to it.
We see that such minima shift as the voltage increases, in an approximately linear fashion. 
A linear fit on the position of the minima with respect to the applied voltage gives a numerical result compatible with a slope value of approximately $1/2\ln{2}$.

\section{Applications}
We will further review the properties of the previous formulae for different setups considering different physical regimes.
While we will review the results for noise and correlations also assuming different electrical biases, we will further discuss the effects of temperature differences.

\subsection{NSN Device}
We now apply the previous results to a very simple example of an NSN system. The two normal terminals are kept at fixed electrochemical potentials
$\mu_1=eV_1$, $\mu_2=eV_2$ and temperatures $T_1$, $T_2$, and transport is mediated by a grounded superconductor ($\mu_s=0$) in the middle.
The study of the behaviour of noise under both electrical \textit{and} thermal biases has already been done before in this kind of systems~\cite{golubevCooperPairSplitting2023}, but here we will analyze the temperature-bias dependence of noise in full detail. We will also make a comparison with the dependence of noise on electrical biases, especially in the shot regimes, which, to our knowledge, has never been done before.
We will consider the low energy limit, well within the subgap regime $E\ll\Delta$ \footnote{Since the superconductor is grounded, the system has three terminals.
}.
We wish indeed to discuss the general form of the current correlations without including strong energy-dependent effects.

\begin{figure}
    \centering
    \includegraphics[width=0.4\textwidth]{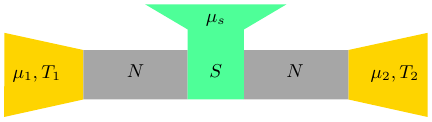}
    \caption{Scheme of the NSN device described in the text, with arbitrary potentials and temperatures at the normal contacts.}
    \label{fig:nsn}
\end{figure}

In this case, the scattering matrix is energy-independent.
Furthermore, we will assume for simplicity that the system is spatially symmetric (i.e. it is symmetric under the exchange of labels $1$ and $2$, see also \Cref{fig:nsn}), and spin degenerate, with only one open transport channel per lead.
In such case, without loss of completeness, we could consider the scattering problem only for, i.e., spin-up electrons and spin-down holes, and also we could parametrize the $ee$ and $he$ scattering matrix blocks as follows:
\begin{equation}
    \begin{gathered}
        \mathbf{s}^{ee} \! = \! e^{i\phi}\begin{pmatrix}
            \sqrt{\mathcal{R}}  &   i\sqrt{\mathcal{T}} \\
            i\sqrt{\mathcal{T}}  &   \sqrt{\mathcal{R}} \\
        \end{pmatrix} \! , \,
        \mathbf{s}^{he} \! = \! e^{i\psi}\begin{pmatrix}
            \sqrt{\mathcal{R}^A}  &   i\sqrt{\mathcal{T}^A} \\
            i\sqrt{\mathcal{T}^A}  &   \sqrt{\mathcal{R}^A} \\
        \end{pmatrix} \! , \\
        \mathcal{R},\mathcal{T},\mathcal{R}^A,\mathcal{T}^A\in[0,1];\;
        \mathcal{R} + \mathcal{T} + \mathcal{R}^A + \mathcal{T}^A = 1;\;
        \phi,\psi\in\mathbb{R},
    \end{gathered}
\end{equation}
The $\mathcal{R},\mathcal{T}$ ($\mathcal{R}^A,\mathcal{T}^A$) parameters are the probabilities of normal (Andreev) reflection and transmission across the device, respectively, while the relative phase between the normal and Andreev processes' types is $\phi-\psi$.
In our spin degenerate and energy-independent case we have the other two blocks $eh$ and $hh$ simply related to the other two by PHS
~\cite{anantramCurrentFluctuationsMesoscopic1996}.
In this case the matrix of transmission functions $(\ell^\pm)_{ij}\equiv\ell_{ij}^{\pm}$ reads
\begin{align}
    \ell^{\pm} &= \begin{pmatrix}
        1-\mathcal{R}\pm\mathcal{R}^A &   -\mathcal{T}\pm\mathcal{T}^A  \\
        -\mathcal{T}\pm\mathcal{T}^A  &   1-\mathcal{R}\pm\mathcal{R}^A \\
    \end{pmatrix}.
    \label{eq:nsntilde}
\end{align}
From these quantities, we can easily compute the local and nonlocal DC transport coefficients, such as the charge and heat conductances, according to the expressions given in \Cref{eq:conductmatr}~\cite{claughtonThermoelectricPropertiesMesoscopic1996}.

\subsubsection{Background noise}
In this simplified regime, the background noise, \Cref{eq:backlinenindell}, is given by
\begin{multline}
    \frac{1}{2k_B G_0}
    \begin{pmatrix}
    \Bar{S}_{11} & \Bar{S}_{12} \\
    \Bar{S}_{21} & \Bar{S}_{22}
    \end{pmatrix}
    \\
    =\begin{pmatrix}
        2 T_{1}\ell^+_{11}
             \! + \! (T_1 \! - \! T_2)\ell^-_{12}   &%
        T_2 \ell^+_{12}
             \! + \!  T_1 \ell^+_{21}   \\
        T_1 \ell^+_{21}
             \! + \!  T_2 \ell^+_{12}   &%
        2 T_{2}\ell^+_{22}
             \! + \! (T_2 \! - \! T_1)\ell^-_{21}   \\
    \end{pmatrix}.
\end{multline}
As we have noticed before, for cross-correlators (off-diagonal terms) the background noise can be thought of as being made up only of a Johnson-Nyquist-like component (which depends only on $\ell^+_{ij}$ functions).
However, in self-correlations, another term, proportional to the temperature differences, appears, which is equivalent in form to a transmitted thermal noise contribution induced by thermal biases, mediated by $\ell^-_{ik}$ transmission functions, which indeed enter in the definition of the \textit{heat} conductance.

\subsubsection{Electrical shot-noise}
Now, we investigate the excess noise's formulae, both in the electrical shot noise limit at thermal equilibrium and, in the next subsection, the thermal shot noise regime at electrical equilibrium.

To investigate the electrical shot noise limit, we assume to keep grounded with the superconductor the drain contact ($eV_{2}=0$) while keeping $eV_{1} \equiv e\Delta V \gg k_B T_i$ in order for the shot regime to emerge.
Under these conditions, we get:
$\int dE D_{1e1h}(E) / 2 
= \int dE D_{1e2e}(E)
\approx \abs{e\Delta V}$
and similarly for the other generalized voltage differences in the other combinations of $k,l,\gamma,\delta$ from \Cref{eq:excess-enind}.
Thus we obtain the self-correlators' excess noise expressions:
\begin{align}
    & \begin{aligned}
        \frac{
            \tilde{S}_{11}}{
            2G_0|e\Delta V|
        } \approx &
        \left[
            \left( 1-\ell^-_{11} \right)
            - \left( 1-\ell^+_{11} \right)^2
        \right] \\
        = & \left[
            \left( \mathcal{R} + \mathcal{R}^A \right)
            - \left( \mathcal{R} - \mathcal{R}^A \right)^2
        \right],
    \end{aligned} 
        \label{eq:electricself1}\\
    & \begin{aligned}
        \frac{
            \tilde{S}_{22}}{
            2G_0|e\Delta V|
        } \approx &
        \left[
            -\ell^-_{21}-\ell^{+2}_{21}
        \right] \\
        = & \left[
            \left( \mathcal{T} + \mathcal{T}^A \right)
            - \left( \mathcal{T} - \mathcal{T}^A \right)^2
        \right],
    \end{aligned} 
        \label{eq:electricself2}
\end{align}
where the second equalities are obtained by directly using the definitions $\ell^\pm$ in \Cref{eq:nsntilde}.
As for the background noise, self-correlators show a mixed dependence on both $\ell^+$ and $\ell^-$, quadratic and linear respectively, as reported elsewhere \cite{akhmerovQuantizedConductanceMajorana2011,denisovChargeneutralNonlocalResponse2021,denisovHeatModeExcitationProximity2022}.
Clearly, from these formulae, it is easy to verify that all the self-correlators are positive, as expected, since $0\leq \mathcal{T},\mathcal{T}^A,\mathcal{R},\mathcal{R}^A\leq 1$.
Remarkably, both self-correlators
are, in general, non-zero even in the \textit{absence} of a net charge current.
In fact, e.g. in \Cref{eq:electricself2}, if $\mathcal{T}=\mathcal{T}^A$ there is a net zero charge current (see \Cref{eq:nsntilde}), but still a non-zero electrical shot noise, actually proportional to $\left( \mathcal{T}+\mathcal{T}^A \right)=\ell^-_{12}$ \cite{bubisThermalConductanceNonequilibrium2021,denisovChargeneutralNonlocalResponse2021,denisovHeatModeExcitationProximity2022}. This represents the specific case where the current of electrons is exactly compensated by the Andreev reflected holes, although the two fluxes contribute independently to the partition noise.
This fact shows the self-correlators' ability
to probe the quasiparticles' transport properties irrespective of their charge.

On the other hand, the cross-correlators' excess noise  is given by
\begin{equation}
        \frac{
            \tilde{S}_{12}}{
            2G_0|e\Delta V|
        } \approx
        \left[
            \left(
                1-\ell^+_{11}
            \right) \ell^+_{21}
        \right]
        =  \left[
            \left( \mathcal{R} - \mathcal{R}^A \right)
            \left( \mathcal{T}^A - \mathcal{T} \right)
        \right],
\end{equation}
and depends only on the \textit{electrical} properties of the device, i.e. on the $\ell^+$ functions (note that, by symmetry, $\tilde{S}_{12}= \tilde{S}_{21}$).

\subsubsection{Thermal shot-noise}
Let us now consider the case where $T_1\gg T_2$ and 
in the absence of voltage biases ($V_{k}=V_s,\forall k$).
By using~\Cref{eq:excess-enind,eq:thermalshot}, for the self-correlators we obtain
\begin{align}
    & \begin{aligned}
        \frac{
            \tilde{S}_{11}}{
            2G_0\eta k_B T_1
        } \approx &
        \left[
            \left( 1-\ell^-_{11} \right)
            - \left( 1-\ell^-_{11} \right)^2
        \right] \\
        = & \left[
            \left( \mathcal{R} + \mathcal{R}^A \right)
            \left( 1-\mathcal{R} - \mathcal{R}^A \right)
        \right],
    \end{aligned} 
        \label{eq:thermalself1} \\
    & \begin{aligned}
        \frac{
            \tilde{S}_{22}}{
            2G_0\eta k_B T_1
        } \approx &
        \left[
            -\ell^-_{21}-\ell^{-2}_{21}
        \right] \\
        = & \left[
            \left( \mathcal{T} + \mathcal{T}^A \right)
            \left( 1-\mathcal{T} - \mathcal{T}^A \right)
        \right],
    \end{aligned} 
        \label{eq:thermalself2}
\end{align}
which, unlike \Cref{eq:electricself1,eq:electricself2}, only depend on the functions $\ell^-$ .
Notice that, when written in terms of $\mathcal{R},\mathcal{R}^A,\mathcal{T},\mathcal{T}^A$, the self-correlators are expressed in the peculiar typical form $x(1-x)$,
i.e., as a partition factor with $x$ being 
the \textit{total}  transmission $(\mathcal{T}+\mathcal{T}^A)$ and \textit{total} reflection $(\mathcal{R}+\mathcal{R}^A)$ probability.
Interestingly, scattering probabilities enter with a plus sign, irrespective of the sign of the charge carriers.
This is consistent with the fact that we are at electrical equilibrium, and only thermal fluctuations are present in the device~\footnote{Note that the energy-independent assumption simplifies the analysis a lot since we can exclude any concurrent thermoelectrical effect.}.
Albeit the superconductor could absorb and emit charges via Andreev processes, it is not able to exchange energy with the rest of the system, and this is shown in the partitioning properties of the charge current's thermal fluctuations.
We stress once more that this property can be clearly observed only in the self-correlators in the thermal shot-noise.

Cross-correlators' excess noise instead has an even more complex structure 
in the presence of thermal biases with respect to the electrical shot-noise case.
Indeed, $\tilde{S}_{12}$ becomes
\begin{equation}
    \frac{
        \tilde{S}_{12}}{
        2G_0\eta k_B T_1
    } \approx 
    \left( \mathcal{R} - \mathcal{R}^A \right)
    \left( \mathcal{T}^A - \mathcal{T} \right) 
    + 4 \sqrt{
        \mathcal{R}
        \mathcal{R}^A
        \mathcal{T}
        \mathcal{T}^A
    },
\end{equation}
where we have an additional interference term: a product of all the transmission and reflection scattering probabilities (normal and Andreev), which can be non-zero only when superconductivity is present.
This is also a term that, in more general cases, could account for the relative phases induced by all different processes that could occur to scattering particles incoming at a system's contact.
Furthermore, it is easy to conclude that without superconductors ($\mathcal{R}^A=\mathcal{T}^A=0$) the ratio between electrical and thermal shot noise would satisfy
\begin{equation}
\frac{\tilde{S}_{12}(\Delta V\gg T_1,T_2)}{\tilde{S}_{12}(T_1\gg\Delta V,T_2)}\approx \frac{e\Delta V}{\eta k_B T_1}.
\end{equation}
Therefore, in the presence of superconductors, for particle energies much smaller than the gap, this universal ratio can generally be different, showing that a comparison between electrical and thermal shot noise could be helpful to provide indirect evidence of the presence of superconducting proximity.

\subsection{Chiral Device}\label{sec:chiral}

\begin{figure}
    \centering
    \includegraphics[width=\linewidth]{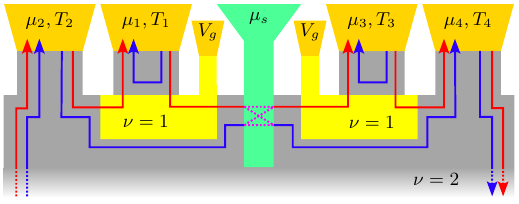}
    \caption{Scheme of the hybrid superconducting-quantum Hall device described in \Cref{sec:chiral}.
    For simplicity we suppose the quantizing magnetic field is directed towards the reader, inducing Zeeman splitting between spin up and down quasiparticle bands.
    Therefore, the red (blue) lines represent the spin up (down) electron- and hole-like edge states.}
    \label{fig:dev-setup}
\end{figure}

As a second example, we study the noise properties of another realistic physical system \cite{giovannettiMultichannelArchitectureElectronic2008,karmakarControlledCouplingSpinResolved2011,karmakarElectronicInterferometerBased2013,chirolliInteractionsElectronicMachZehnder2013,karmakarNanoscaleMachZehnderInterferometer2015,leeInducingSuperconductingCorrelation2017,beconciniNonlocalSuperconductingCorrelations2018}, consisting of an integer quantum Hall bar at filling factor $\nu=2$ with a superconducting scatterer along the edge states' path, see \Cref{fig:dev-setup}.
We assume that the two edge states are spin-resolved as a consequence of the Zeeman splitting caused by the quantizing magnetic field (applied perpendicularly to the bar's plane).
Spin-up electrons are represented as red lines, and spin-down electrons as blue lines. Source (drain) contacts are represented as dark yellow regions on the left (right), characterized by chemical potentials $\mu_1$ and $\mu_2$ ($\mu_3$ and $\mu_4$) and temperatures $T_1$ and $T_2$ ($T_3$ and $T_4$)~\footnote{The possibility to apply different electronic temperatures to the contacts can be more easily realized at very low temperatures where electronic degrees of freedom decouple from the phononic degrees of freedom.}.
Light yellow rectangles represent gates which control the filling factor in the regions beneath them, used to locally deplete one spin band and thus create $\nu=1$ regions.

The two edge states are contacted separately, both to the source and drain contacts, using a four-terminal setup.
For simplicity, we assume that no mixing occurs between co-propagating edge states in the normal regions.
We also neglect any possible effect of the electron-electron interaction between the two channels~\cite{braggioNonlocalThermoelectricDetection2024}.
All relevant scattering processes between the co-propagating modes happen at the scattering region between the gates (green finger in Fig.~\ref{fig:dev-setup}), which can be superconducting or a simple spin-mixing insertion.
In other words, the spin-up (spin-down) quasiparticles coming from contact 1 (2) running along the outer (internal) edge are coupled by the finger and are collected at contacts 3 (spin up) and 4 (spin down).
If the finger is a conventional BCS singlet superconductor, electrons propagating on one edge state are converted into holes on the other edge via an Andreev transmission process~\cite{clarkeExoticCircuitElements2014,houCrossedAndreevEffects2016,leeInducingSuperconductingCorrelation2017}. If instead the finger is a spin-mixer, it converts electrons in one edge state into electrons in the other edge changing the electron spin. 

The scattering matrix of the device has dimension $16\times 16$, accounting for 4 contacts and the spin degree of freedom.
Many elements of such matrix are, however, either 0 or 1 because of the chiral nature of transport in the device. For example, the reflection of spin-down electrons in lead 1 occurs with probability 1, while the reflection of spin-up occurs with probability 0.
Therefore, for simplicity, we organize the non-trivial  scattering elements into the following matrix
\vspace{-7pt}
\begin{equation}
    \mathbf{t}(E) =
    \begin{NiceMatrixBlock}
        \begin{pNiceArray}%
        {%
            cc:%
            cc%
        }
            s^{ee}_{31}(E)  &
                s^{ee}_{32}(E)  &
                s^{eh}_{31}(E)  &
                s^{eh}_{31}(E)  \\
            s^{ee}_{41}(E)  &
                s^{ee}_{42}(E)  &
                s^{eh}_{41}(E)  &
                s^{eh}_{41}(E)  \\
            \hdottedline
            s^{he}_{31}(E)  &
                s^{he}_{32}(E)  &
                s^{hh}_{31}(E)  &
                s^{hh}_{31}(E)  \\
            s^{he}_{41}(E)  &
                s^{he}_{42}(E)  &
                s^{hh}_{41}(E)  &
                s^{hh}_{41}(E)
        \end{pNiceArray}
    \end{NiceMatrixBlock},
    \label{eq:transfer-chiral}
\end{equation}
where, again, the various blocks are related by PHS, although now the system is not spin degenerate as in the previous NSN case~\cite{jacquodOnsagerRelationsCoupled2012}.
The matrix $\ell^{\pm}(E)$ can be written accordingly (using the full definition of the scattering matrix) as
\vspace{-8pt}
\begin{equation}
    \ell^{\pm}(E)
    = \begin{pmatrix}
        1                   &   -1                  &   0   &   0   \\
        0                   &   2                   &   0   &   -2  \\
        \ell^{\pm}_{31}(E)  &   \ell^{\pm}_{32}(E)  &   1   &   0   \\
        \ell^{\pm}_{41}(E)  &   \ell^{\pm}_{42}(E)  &   -1  &   2
    \end{pmatrix}.
\end{equation}
We emphasize the fact that the matrix is not symmetric as a result of the chiral nature of transport in the system,
and its structure shows again that the only non-trivial transmission processes are the ones describing the scattering from contacts 1 and 2 to contacts 3 and 4.

Once the transmission functions are known, the background noises can be easily calculated using the formalism developed above, obtaining
\begin{widetext}
    \begin{equation}
        \renewcommand{\arraystretch}{1.7}
        \frac{\Bar{S}}{2 k_B G_0}
        \! = \! \begin{NiceMatrixBlock}
            \begin{pNiceArray}{
                Wc{30mm}
                Wc{30mm}
                Wc{40mm}
                Wc{50mm}
            }
                T_1 \! + \! T_2   &
                    -T_2  &
                    \int \! \! \frac{dE}{2k_B}
                        \! F_{1e}(E) \ell^+_{31}(E)    &
                    \int \! \! \frac{dE}{2k_B}
                        \! F_{1e}(E) \ell^+_{41}(E)    \\
                -T_2  &
                    2 T_2 \! + \! 2 T_4   &
                    \int \! \! \frac{dE}{2k_B}
                        \! F_{2e}(E) \ell^+_{32}(E)    &
                    \int \! \! \frac{dE}{2k_B}
                        \! F_{2e}(E) \ell^+_{42}(E)    \\
                \int \! \! \frac{dE}{2k_B}
                    \! F_{1e}(E) \ell^+_{31}(E)    &
                    \int \! \! \frac{dE}{2k_B}
                        \! F_{2e}(E) \ell^+_{32}(E)    &
                    T_3 \! - \! \sum\limits_{k=1}^2 \!
                        \int \! \! \frac{dE}{2k_B}
                        \! F_{ke}(E) \ell^-_{3k}(E)    &
                    -T_3  \\
                \int \! \! \frac{dE}{2k_B}
                    \! F_{1e}(E) \ell^+_{41}(E)    &
                    \int \! \! \frac{dE}{2k_B}
                        \! F_{2e}(E) \ell^+_{42}(E)    &
                    -T_3  &
                    2 T_4 \! + \! T_3 \! - \! \sum\limits_{k=1}^2 \!
                        \int \! \! \frac{dE}{2k_B}
                        \! F_{ke}(E) \ell^-_{4k}(E)
            \end{pNiceArray}.
        \end{NiceMatrixBlock}
    \end{equation}
\end{widetext}
We notice the usual structure of Johnson-Nyquist noises mediated by $\ell^+_{ij}(E)$ and $\ell^-_{ij}(E)$ functions in cross- and self-correlators, respectively.
However, for this chiral case, we manage to isolate the dependence on $\ell^-(E)$ thanks to the peculiar geometry of the device.

In this system, the excess noises have a non-trivial form 
\begin{equation}
    \tilde{S} =
    \begin{pmatrix}
        0   &   0   &   0   &   0   \\
        0   &   0   &   0   &   0   \\
        0   &   0   &   \tilde{S}_{33}   &   \tilde{S}_{34}   \\
        0   &   0   &   \tilde{S}_{43}   &   \tilde{S}_{44}
    \end{pmatrix},
\end{equation}
that is, we have excess noise contributions only at or between contacts 3 and 4.
This happens because the partition noise is determined by the scattering processes occurring at the superconducting finger on the edge states that come from contacts 1 and 2.
Therefore, only noise \textit{downstream} to the scatterer (in the sense of the edge states' propagation direction) can contain contributions due to current \textit{partitioning}.
In addition, these contributions' magnitude can depend only on the squared population differences between quasiparticles that come from contacts \textit{upstream} to the scatterer. At the same time, they cannot be affected at all neither by the voltages nor by the temperatures of the downstream terminals (3 and 4), i.e. they can depend only on $D_{1\gamma 2\delta}(E)$ (with $\gamma,\delta\in\{e,h\}$).

\subsubsection{Thermal shot regime}
The excess noise at contacts $3$ and $4$ in the thermal shot regime ($T_1\gg T_2$) at $\mu_1=\mu_2=\mu_s=0$, in general, can be written only in terms of $D_{1e2e}(E)$ as \footnote{Indeed, $D_{1e2h}(E)$ are equal to the $D_{1e2e}(E)$ ones since, at electrical equilibrium, particles and holes' Fermi distributions are equal. Alternatively, the two types behave in the same way under the action of temperature gradients.}
\begin{multline}
    \frac{\tilde{S}_{ij}}{2G_0}
    = \int dE D_{1e2e}(E) \Big\llbracket
        \delta_{ij} 
        \left(
            N_i \delta_{i1} - \ell^-_{i1}(E)
        \right) \\
        - \Tr\left[
            \left(
                s^{ee\dagger}_{i1}(E)s^{ee}_{i1}(E)
                -s^{he\dagger}_{i1}(E)s^{he}_{i1}(E)
            \right)
            \right. \\ \times \left.
            \left(
                s^{ee\dagger}_{j1}(E)s^{ee}_{j1}(E)
                -s^{he\dagger}_{j1}(E)s^{he}_{j1}(E)
            \right)
        \right] \\
        - 2\Re\left\{
            \Tr\left[
                s^{ee\dagger}_{i1}(E)s^{eh}_{i1}(E)
                \right. \right. \\ \times \left. \left.
                \left(
                    s^{eh\dagger}_{j1}(E)s^{ee}_{j1}(E)
                    -s^{hh\dagger}_{j1}(E)s^{he}_{j1}(E)
                \right)
            \right]
        \right\}
    \Big\rrbracket.
\end{multline}
By substituting the actual expression of the scattering matrix for our chiral system, assuming again energy independence for simplicity, we get the following results for self-correlators
\begin{equation}
        \frac{\tilde{S}_{ii}}{2G_0\eta k_B T_1}
            = \Big\llbracket
                - \ell^-_{i1}
                \left[
                    1 + \ell^-_{i1}
                \right]
            \Big\rrbracket,
            \end{equation}
and for cross-correlators
\begin{equation}
    \frac{\tilde{S}_{34}}{2G_0\eta k_B T_1}
        = \Big\llbracket
            - \ell^+_{31}\ell^+_{41}
        \Big\rrbracket.
\end{equation}
The self-correlations show, as expected, a marked dependence on the matrix $\ell^-$, but we notice a difference in the cross-correlations' result with respect to the NSN case: here there is no additional interference term $\propto \sqrt{\mathcal{R}
        \mathcal{R}^A
        \mathcal{T}
        \mathcal{T}^A
}$ anymore.
Indeed, such a term would be present only if contact 3 (4) could collect also spin down (up) particles coming from the scattering finger.
In this setup, though, this is not possible since 
each contact collects only 2 possible spin-polarized channels coming from the scattering region, and thus, cross-correlations' excess noise $\tilde{S}_{34}$ cannot be sensitive to the interference between the partitioning processes involving the same spin.

\subsubsection{General excess noise behaviour}\label{sec:deltatsmar}
In the general case (where the values of $V_1$, $V_2$, $T_1$ and $T_2$ are arbitrary)
all terms of the excess noises appearing in \Cref{eq:excess} should be considered.
It is useful to parametrize the matrix (\ref{eq:transfer-chiral})~\cite{ditaParametrisationUnitaryMatrices1982,ditaParametrisationUnitaryMatrices1994}, which describes the superconducting finger, in such a way to account for various effects, such as the presence of magnetic impurities.
This can be done by using
a matrix which results from the composition 
of two sequential processes describing two different phenomena: Andreev reflection (AR) and spin mixing (SM).
For simplicity, we will also assume that the scattering amplitudes are energy-independent.
AR processes are described by the matrix
\begin{equation}
    \begin{gathered}
        \begingroup
        \setlength\arraycolsep{1pt}
            \mathbf{t}_{\rm{AR}} 
            \!
            = 
            \! \! \!
            \begin{pNiceArray}%
            {cc:cc}
                e^{i\psi} \! \sqrt{a}e^{i\alpha}  &
                    0   &
                    0   &
                    -e^{i\psi} \! \sqrt{b}e^{-i\beta}  \\
                0   &
                    e^{-i\psi} \! \sqrt{a}e^{i\alpha}  &
                    e^{-i\psi} \! \sqrt{b}e^{-i\beta}  &
                    0   \\
                \hdottedline
                0   &
                    -e^{-i\psi} \! \sqrt{b}e^{i\beta}  &
                    e^{-i\psi} \! \sqrt{a}e^{-i\alpha}  &
                    0   \\
                e^{i\psi} \! \sqrt{b}e^{i\beta}  &
                    0   &
                    0   &
                    e^{i\psi} \! \sqrt{a}e^{-i\alpha}
            \end{pNiceArray}
            \!
            ,
        \endgroup \\
        a,b\in [0,1];\, a+b=1;\, \psi,\alpha,\beta\in\mathbb{R},
    \end{gathered}
\end{equation}
where $a$ and $b$ are the probabilities of normal and Andreev processes, respectively.
SM processes, instead, are described by
\begin{equation}
    \begin{gathered}
        \begingroup
        \setlength\arraycolsep{1pt}
            \mathbf{t}_{\rm{SM}} 
            \!
            = 
            \! \! \!
            \begin{pNiceArray}%
            {cc:cc}
                e^{i\phi} \! \sqrt{r}e^{i\rho} & e^{i\phi} \! \sqrt{t}e^{-i\theta} & 0 & 0 \\
                e^{i\phi} \! \sqrt{t}e^{i\theta} & -e^{i\phi} \! \sqrt{r}e^{-i\rho} & 0 & 0 \\
                \hdottedline
                0 & 0 & e^{-i\phi} \! \sqrt{r}e^{-i\rho} & e^{-i\phi} \! \sqrt{t}e^{i\theta} \\
                0 & 0 & e^{-i\phi} \! \sqrt{t}e^{-i\theta} & -e^{-i\phi} \! \sqrt{r}e^{i\rho} \\
            \end{pNiceArray}
            \!
            ,
        \endgroup \\
        r,t\in [0,1];\, r+t=1;\, \phi,\rho,\theta\in\mathbb{R},
    \end{gathered}
\end{equation}
with $t$ and $r$ being the probabilities of transmission with and without mixing, respectively.
The matrix describing the composition of the two processes is then given by the product of the two, namely $\mathbf{t}=\mathbf{t}_{\rm SM}\cdot\mathbf{t}_{\rm AR}$.

By using \Cref{eq:excess} we obtain the  excess noise for the self-correlators
    \begin{multline}
        \frac{\tilde{S}_{33}}{2G_0}
        = \frac{\tilde{S}_{44}}{2G_0}
        = + rt(1 - 4ab) \int dE D_{1e2e}(E) \\
        + ab \int dE D_{1e2h}(E)
        + 2rtab
        \bigg[
            - 2 k_B T_1
            - 2 k_B T_2
        \\
        \left.
            + e(
                \Delta V_{1e2h} 
                + \Delta V_{1e2e}
            )
            \coth(
                \frac{
                    e(
                        \Delta V_{1e2h} 
                        + \Delta V_{1e2e}
                    )
                }{
                    2k_BT_1
                }
            )
        \right.
        \\
        \left.
            + e(
                \Delta V_{1e2h} 
                - \Delta V_{1e2e}
            )
            \coth(
                \frac{
                    e(
                        \Delta V_{1e2h}
                        - \Delta V_{1e2e}
                    )
                }{
                    2k_BT_2
                }
            )
        \right] ,
        \label{eq:two-exc-smar-self}
    \end{multline}
and for the cross-correlators
    \begin{multline}
        \frac{\tilde{S}_{34}}{2G_0}
        = \frac{\tilde{S}_{43}}{2G_0}
        = - rt(1 - 4ab) \int dE D_{1e2e}(E) \\
        + ab \int dE D_{1e2h}(E)
        - 2rtab
        \bigg[
            - 2 k_B T_1
            - 2 k_B T_2
        \\
        \left.
            + e(
                \Delta V_{1e2h} 
                + \Delta V_{1e2e}
            )
            \coth(
                \frac{
                    e(
                        \Delta V_{1e2h} 
                        + \Delta V_{1e2e}
                    )
                }{
                    2k_BT_1
                }
            )
        \right.
        \\
        \left.
            + e(
                \Delta V_{1e2h} 
                - \Delta V_{1e2e}
            )
            \coth(
                \frac{
                    e(
                        \Delta V_{1e2h}
                        - \Delta V_{1e2e}
                    )
                }{
                    2k_BT_2
                }
            )
        \right]\,.
        \label{eq:two-exc-smar}
    \end{multline}
Here we have used the two voltage variables $\Delta V_{1e2e} = V_1-V_2$ and $\Delta V_{1e2h} = V_1+V_2$. 
We have made this choice since the first terms in both equations depend only on $\Delta V_{1e2e}$, while the second terms depend only on $\Delta V_{1e2h}$, even if the two temperatures $T_{1}$ and $T_2$ are different.
Notice that the second terms are the leading ones when Andreev processes dominate over spin-mixing processes ($ab\gg rt$), while the first terms are leading when spin-mixing processes dominate over Andreev ones ($rt\gg ab$).
Indeed, the partition factor $ab$ ($rt$) determines the excess noise contributions due to AR (SM).
Interestingly, these two contributions enter the excess noise of the cross-correlators with a different sign:
the sign in front of the $rt$-dependent contribution is negative, while $ab$-dependent contribution is positive.
This is a result of the fact that, when undergoing an SM (AR) process, an excitation keeps (changes) its charge.

\begin{figure}
    \subfloat[\label{fig:chiral-deltat-zero}]{%
        \includegraphics[width=0.48\textwidth]{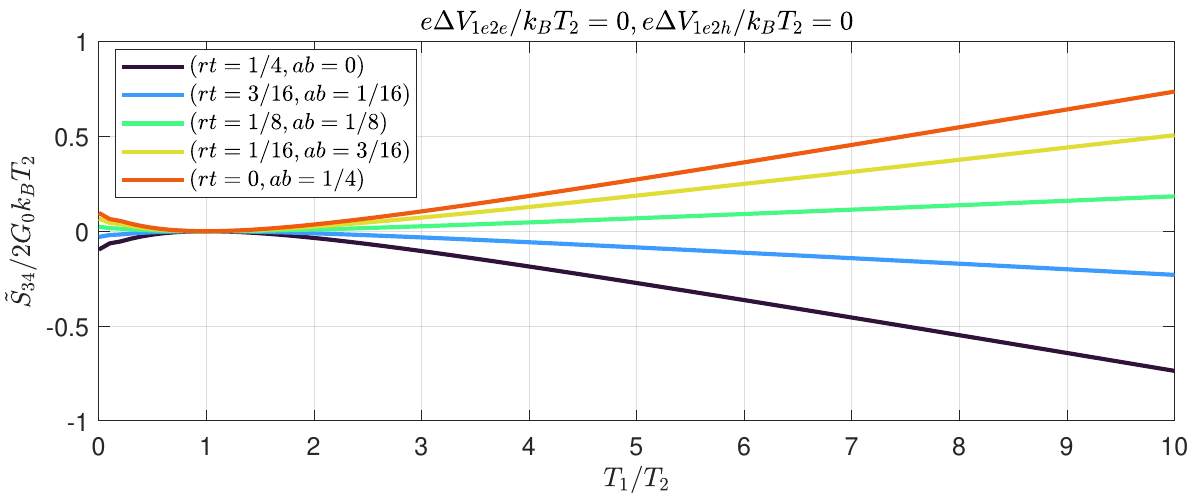}%
    }
    
    \subfloat[\label{fig:chiral-deltat-ten}]{%
        \includegraphics[width=0.47\textwidth]{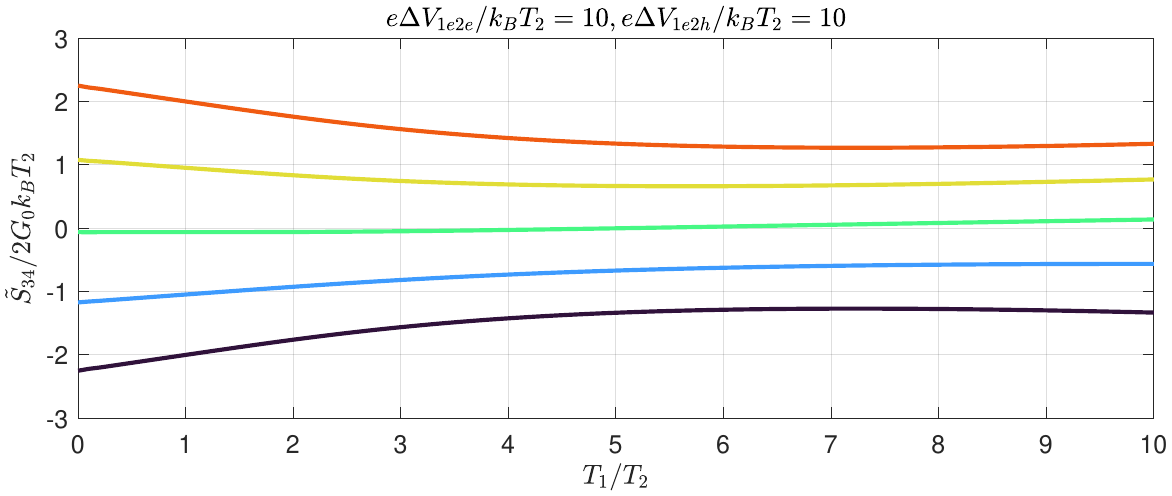}%
    }
    \caption{Excess noise for the cross-correlator $\tilde{S}_{34}$ of the system in \Cref{sec:deltatsmar}, for various values of the partition factors $ab$ and $rt$  plotted against the  temperature ratio $T_1/T_2$.
    The upper panel corresponds to the noise curves obtained when no voltage biases are imposed between contacts 1 and 2, while the lower panel is obtained when $\Delta V_{1e2e}=\Delta V_{1e2h}=10 k_BT_2$
    The various curves in each panel correspond to different combinations of values of the partition factors, shown in the legend in the upper panel.
    }
    \label{fig:chiral-smar-thermal-shot}
\end{figure}
Together with the relative magnitude of the two partition factors $ab$ and $rt$, also the absolute values of the voltage variables ($\Delta V_{1e2e}$ and $\Delta V_{1e2h}$) determine the behaviour of 
the excess noise when the temperature's ratio $T_1/T_2$ is varied.
In \Cref{fig:chiral-smar-thermal-shot}, for example, we plot the cross-correlator's excess noise as a function of the temperatures' ratio and for different values of $rt$ and $ab$. 
The curves with colours going from black to red correspond to increasing values of $ab$ and decreasing values of $rt$.
Both panels show that the thermal shot noise is positive when $ab$ is greater than $rt$.
However, when the two values are close or equal, the sign of the thermal shot noise actually depends on the specific case we are analyzing~\footnote{For instance, if we define $\mathbf{t}=\mathbf{t}_{\rm{AR}}\cdot\mathbf{t}_{\rm{SM}}$, we obtain a negative thermal shot noise if $ab=rt$.}.
The sign change is due to the fact that the two processes have opposite effects on the charge currents' correlations since the SM (AR) \textit{anticorrelates} (\textit{correlates}) them.

In \Cref{fig:chiral-deltat-zero} we fix $\Delta V_{1e2e}=\Delta V_{1e2h}=0$: in the absence of voltage biases, all the curves touch zero at $T_1=T_2$.
However, for finite voltage biases such as in \Cref{fig:chiral-deltat-ten}, where we fix $\Delta V_{1e2e}=\Delta V_{1e2h}=10k_BT_2$, the excess noise acquires a voltage floor: since $\Delta V_{1e2h}\neq 0$, when $ab>rt$ the minimum of the noise is positive, but near $ab=rt$ it gets to zero (the exact condition is, again, system specific), and it changes sign when $ab<rt$, provided that also $\Delta V_{1e2e}\neq 0$.
In conclusion, we see that the sign of the voltage floor as a function of the generalized voltage difference of the terminals will indicate which process
contributes the most to noise, between SM and AR.
Thus, we find again that the analysis of the behaviour of the excess noise while varying the temperature could potentially return intriguing information about the intrinsic properties of hybrid systems.

\section{Conclusions}

We detailed and analyzed the properties of the charge current noise in hybrid normal-superconducting devices, where the superconductors are grounded and kept under fully out-of-equilibrium conditions. Indeed, we consider not only the voltage biases but, more intriguingly, also the temperature differences.
We found that thermally out-of-equilibrium noise strongly differs from the one without superconducting correlations, both in the background and in the excess (partition) noise contributions.
In particular, we found that the self-correlations' background noise, in the presence of temperature differences,
is not only dictated by the charge transmission functions $\ell^+$, a quantity related to the charge transport (electrical conductance), but also by $\ell^-$, which is instead related to the heat transport (thermal conductance).
As a consequence, this effect could be exploited to gain information about heat transport
via just charge current noise measurements only instead of measuring charge and heat transport separately. This may be quite advantageous since it is not usually so simple to directly measure heat currents.
Furthermore, we explored the generic behaviour of the excess noise when both voltage and thermal biases are present.
When superconductors are grounded, the partition noise depends not only on the bias differences but
also on their sum, as a direct result of Andreev's processes.
We then analyzed the behaviour of the excess noise in the presence of thermal biases and identified the thermal shot noise regime, both in the low- and high-temperature limits.
This result generalizes to hybrid systems the results obtained for purely normal systems \cite{sivreElectronicHeatFlow2019,larocqueShotNoiseTemperatureBiased2020}.

Finally, we have
considered two experimentally relevant examples: a normal-superconductor-normal (NSN) system and the proximitized edge states of a Quantum Hall bar.
In addition to the background noise behaviour, we focused on the electrical and thermal shot noise regimes, highlighting their differences.
What we noticed is the fact that, as we also remarked for the background noise, highly non-trivial dependencies on the $\ell^-$ functions appear in the excess noises of self-correlators, and these are even stronger for the thermally-biased case with respect to the voltage bias case.
The excess noise in the electrical shot limit for the cross-correlators, instead, shows simply a dependence on $\ell^+$ functions both in the normal and in the hybrid case, while in the thermal shot-noise regime the two cases differ by an interference term that strictly depends uniquely on the presence of Andreev processes.
Moreover, in the Quantum Hall bar example, we inspected the behaviour outside the thermal shot regime in a simplified energy-independent case, assuming the scattering region to act as a sequence of two processes, Andreev conversion and spin-mixing.
In such a case, the noise depends on four quantities: the difference and the sum of the applied input voltages, as well as the two partition factors of the two scatterers.
The relative magnitude of the two governs the overall behavior of the noise, including the thermal shot limit.

We expect our analysis to prove useful in further investigations of hybrid systems for future quantum technological applications, which rely on the unique and remarkable properties of superconductors under thermal out-of-equilibrium conditions.

\begin{acknowledgments}

We all acknowledge the discussions with Prof. V. Giovannetti and Prof. A Romito. We acknowledge funding from MUR-PRIN 2022 - Grant No. 2022B9P8LN
- (PE3)-Project NEThEQS “Non-equilibrium coherent
thermal effects in quantum systems” in PNRR Mission 4 -
Component 2 - Investment 1.1 “Fondo per il Programma
Nazionale di Ricerca e Progetti di Rilevante Interesse
Nazionale (PRIN)” funded by the European Union - Next
Generation EU and the PNRR MUR project
PE0000023-NQSTI.
AB and FT acknowledge the Royal Society through the International Exchanges between the UK and Italy (Grants No.
IEC R2 192166).
AB acknowledges the EU’s
Horizon 2020 Research and Innovation Framework Programme under Grant No. 964398 (SUPERGATE), No.
101057977 (SPECTRUM) and CNR project QTHERMONANO.

\end{acknowledgments}

\nocite{*}
\bibliography{bibliography}

\end{document}